\newif\iffulledition 
\fulleditiontrue 

\documentclass[runningheads]{llncs}
\usepackage{graphicx}
\usepackage{booktabs}
\usepackage{tabularx}

\begin{document}

\title{Visualising Personal Data Flows: Insights from a Case Study of Booking.com\thanks{This is the full edition of a paper published in Intelligent Information Systems: CAiSE Forum 2023, Zaragoza, Spain, June 12–16, 2023, Proceedings, Lecture Notes in Business Information Processing (LNBIP), Volume 477, pp. 52-60, 2023, Springer Nature, https://doi.org/10.1007/978-3-031-34674-3\_7\}}}

\titlerunning{Visualising personal data flows through privacy policy analysis: Booking.com}

\author{Haiyue Yuan\inst{1}\orcidID{0000-0001-6084-6719}
\and
Matthew Boakes\inst{1}\orcidID{0000-0002-9377-6240}
\and
Xiao Ma\inst{2}
\and
Dongmei Cao\inst{2}\orcidID{0000-0002-2614-3726}
\and
Shujun Li\inst{1}\orcidID{0000-0001-5628-7328} }
\authorrunning{Yuan et al.}
%
\institute{Institute of Cyber Security for Society (iCSS) \& School of Computing, University of Kent, United Kingdom\\
\email{\{h.yuan-211, m.j.boakes, s.j.li\}@kent.ac.uk}
\and
Nottingham Business School, Nottingham Trent University, United Kingdom\\
\email{\{xiao.ma, dongmei.cao\}@ntu.ac.uk}\\
}

\maketitle

\begin{abstract}
Commercial organisations are holding and processing an ever-increasing amount of personal data. Policies and laws are continually changing to require these companies to be more transparent regarding the collection, storage, processing and sharing of this data. This paper reports our work of taking Booking.com as a case study to visualise personal data flows extracted from their privacy policy. By showcasing how the company shares its consumers' personal data, we raise questions and extend discussions on the challenges and limitations of using privacy policies to inform online users about the true scale and the landscape of personal data flows. This case study can inform us about future research on more data flow-oriented privacy policy analysis and on the construction of a more comprehensive ontology on personal data flows in complicated business ecosystems.
\end{abstract}

\begin{keywords}
Personal data, Data flow, Privacy policy, Data sharing, Travel
\end{keywords}

\section{Introduction}\label{sec:introduction}

Despite the existence of information security policies and data protection laws such as the EU's GDPR (General Data Protection Regulation), over-collection and breaches of personal data are constantly happening in the online world. Such data privacy and security issues are partly due to the complex nature of data collection, processing and sharing processes, where multiple parties are involved and the data owners (more formally called ``data subjects'') often have no clear view of how their personal data flow between different entities. When a data owner uses their social media (e.g., Facebook) account to access an online service, there will be further personal data flows between the social media company (e.g., Meta) and the service provider. 

Many researchers have studied data security and privacy in different scenarios, e.g., security threats in device-to-device (D2D) networks~\cite{Jin-B2018}, privacy issues in data mining operations~\cite{Binjubeir-M2020}, and data security problems in smart homes~\cite{She-W2019}. Many solutions have also been proposed to address such issues, e.g., Razak et al.~\cite{Razak-S2020} proposed a biometric authentication solution so the data holders can better manage data privacy and security in the wireless network and cloud computing storage. Regarding privacy issues relevant to data sharing and processing, Biennier et al.~\cite{Biennier-F2005} proposed to couple a generic authorisation workflow to the business process and to report all the actions on the shared information systems; such a multi-level architecture seemingly provides both security and personal data privacy in a controllable situation. However, the biggest data privacy and security threat caused by the complex situation of personal data sharing among multiple parties~\cite{Such-JM2016} is not sufficiently studied in the literature. 

A recent study~\cite{ioannou2021s} investigated such complexity in the tourism domain, and suggested that, while collecting and using personal data can result in more appealing tourism offers and more efficient travel, it can also lead to security risks and privacy concerns, thereby discouraging some travellers from sharing their personal data with service providers. To that end, it will help if travellers are made aware of what personal data will be collected and shared with whom for what purposes. A common approach is to present a privacy policy to users, and some past studies looked into privacy policies in different perspectives such as their impact on users' privacy perception, attitude and behaviour~\cite{bracamonte2020effects,ibdah2021should}, automate privacy policy analysis~\cite{Andow-B2019,Harkous-H2018,Carlsson-R2022}, and readability and visualisation~\cite{reinhardt2021visual,Carlsson-R2022,Andow-B2019}. However, the current approaches are fragmented without comprehensively addressing the full scale of personal data collection and sharing activities and data flows generated by such activities. Having this in mind, the following research questions are raised in this study. 

\begin{itemize}
\item Can we extract and visualise personal data flows between data owners and different data-consuming parties from the analysis of a privacy policy?

\item Can the visualisation of personal data flows inform the user about what data can be collected, what data can be shared, and the purposes of the data collection and sharing?
\end{itemize}  

To answer these research questions, this study aims to advance the existing knowledge on data privacy and security issue, with a focus on understanding how online users' personal data are shared and flow to different data-consuming organisations. By utilising an in-depth case study of Booking.com, this paper reports our work of a privacy policy analysis to obtain more in-depth insights with a visual aid to understand how Booking.com and other associated organisations collect and use personal data from customers of Booking.com. We consider the following four main contributions of this paper.

\begin{itemize}
\item We propose an approach to systematically analysing and reconstructing personal data flows declared in a privacy policy.

\item We report insights about personal data flows derived from privacy policies via a simple data visualisation approach.

\item We have identified the needs to have a more in-depth investigation of privacy policies from other relevant organisations to get a more comprehensive understanding of personal data flows, and lessons learnt from this case study and future research directions.
\end{itemize}

We would like to highlight that this case study is preliminary work: 1) to provide an in-depth understanding of privacy policy analysis to scale up the study further; 2) to facilitate automation of privacy policy analysis as part of our future work; and 3) to establish the basis for constructing more comprehensive graphical models and ontologies as part of our future work.

The rest of this paper is organised as follows. Section~\ref{sec:literature-review-and-concepts} presents recent related work and identified research gaps, followed by presenting the methodology of this study in Section~\ref{sec:methodology}. Then Section~\ref{sec:results} details findings of how Booking.com can collect and share customer data. Section~\ref{sec:discussion-and-conclusion} concludes this paper with more discussion on limitations and future work.

\section{Related Work}
\label{sec:literature-review-and-concepts}

The law in many countries, such as the GDPR in EU member states and the UK, requires service providers to supply privacy policies when collecting personal data is involved and to present such privacy policies in a concise, transparent, intelligible, and easily accessible form~\cite{gdpr_writing_2018}. In addition, service providers increasingly gather more personal data from various sources, including personal mobile devices, by `enhancing' the quality of service to their users. However, knowing what personal data is out there and how to control it remains a challenge for end users. Providing this power could allow end users to control and monetise their data.

Researchers have studied the relationship between the content of online privacy policy and consumer trust and the moderating effect of cultural background. Wu et al.~\cite{wu2012effect} studied the impact of online privacy policies on consumer privacy concerns and confidence, and suggested that online privacy becomes a primary concern because the Internet has no borders or regulated restrictions. As a result, companies collect and use vast quantities of customers' personal information. The authors also indicated that creating a privacy policy is one way to increase consumers' trust and reduce privacy concerns. The researchers found significant relationships between the content of the privacy policy and privacy concern/trust, as well as a willingness to provide personal information and privacy concern/trust. 

In addition, the format of the privacy policy may have an impact on customers' decision to disclose their data. Along this line of research, Bracamonte et al.~\cite{bracamonte2020effects} performed an experimental study which found that users prefer an automated tool for summarising privacy policies when the summary includes a policy excerpt to justify the outcome. Privacy policy summarisation tools are automated applications utilising various machine learning and natural language processing techniques. These tools may solve the problem of scale in the analysis of privacy policies, but they introduce a new obstacle: users are concerned about their reliability and precision. 

Furthermore, regulations like the GDPR have spurred recent efforts to make privacy policies more comprehensible to users. Bracamonte et al.~\cite{bracamonte2020effects} experimentally evaluated the effects of explanatory information on the perception of the results of an automated privacy policy tool. The results indicate that justification information increases behavioural intention and perceptions of credibility and utility. The researchers argued that additional research is necessary to determine what information to present to users to improve the effectiveness and perception of these automated privacy tools~\cite{bracamonte2020effects}. Researchers also looked into the visual design's role in framing and controlling privacy notice interaction~\cite{kitkowska2020enhancing}. Kitkowska et al.~\cite{kitkowska2020enhancing} conducted an online experiment and revealed that people feel powerless when acknowledging notices and have no choice but to agree based on the responses of 88 participants. Their findings indicate that designs that provide control and curiosity may increase valence. In addition, the authors investigated how this influenced their disclosure intent, privacy comprehension, and emotional state. At lower values of users' affective state, there was a significant relationship between trust and privacy concerns. Ibdah et al.~\cite{ibdah2021should} studied users' attitudes and opinions regarding privacy rules, as well as the impact of technical jargon on reading such documents. They questioned whether users comprehend the ramifications of subscribing to the terms and conditions outlined in the privacy policy. Additionally, they investigated the effect of these terms on users' continued use of the service once they are aware of the effects and whether this would result in a change in behaviour for better privacy policies. They found that 77\% of the study's participants acknowledged having read privacy policies. The overwhelming majority of respondents (72.4\%) responded that concerns about (untrusted) service providers are the primary motivation for users to read such rules. Overall they discovered a high association between privacy policy knowledge and users' willingness to disclose their information with online service providers. They also argued that consumers of products or services often overlook privacy policies, which provide insights into the relevant organisations' use of personal data~\cite{ibdah2021should}. 

Moreover, to improve the readability of privacy policies, Reinhardt et al.~\cite{reinhardt2021visual} proposed developing interactive privacy policies based on the concept of nutrition labels. They indicated that online privacy is a prevalent issue and that regulations, court rulings, and changes in business practice have all been altering online users' experience. Harkous et al.~\cite{Harkous-H2018} proposed an automated framework for privacy policy analysis (Polisis) based on deep learning. Based on Polisis, they built an application PriBot to allow users to submit free-form queries on privacy policies to provide better readability of privacy policy. Similarly, Andow et al.~\cite{Andow-B2019} developed PolicyLint, a framework that can automatically generate ontologies from privacy policies through the use of sentence-level natural language processing. By deploying PolicyLint to analyse privacy policies of 11,430 mobile apps, they identified that 14.2\% of the privacy policies contain contradicting content. Carlsson et al.~\cite{Carlsson-R2022} also investigated mobile apps' privacy policies. Differently, they analysed the privacy policies of 32 public sector mobile apps to show the actual network traffic in order to identify inconsistencies between the actual data flows and the declarations of personal data collection and sharing in the privacy policies. Similarly, Jin et al.~\cite{Jin-H2018} investigated data flows from 15,000 Android apps with 6.3 million network traces to 12,000 domains. A dedicated website\footnote{\url{http://android-network-tracing.herokuapp.com/}} was developed to publish visualised details of personal data collection and sharing activities of the apps. 

In the tourism domain, Tussyadiah et al.~\cite{tussyadiah2019privacy} provided a comprehensive review of the research on privacy concerns and tourism-related behaviours. In light of existing and emerging privacy issues in tourism, they reviewed existing research on privacy concerns, cognitive biases in privacy decisions, and privacy nudges. By examining the current state of research on information privacy from multiple disciplinary perspectives, they recommended research priorities based on emerging issues in tourism to ensure that tourists make more informed decisions regarding the disclosure of personal information related to their travels. Similarly, Ioannou et al.~\cite{ioannou2021s} conducted a survey to study the relationship between travellers' privacy concerns and online data disclosures. They found that disclosures of personal data could be exchanged for tourism offers to enhance the travellers' overall travel experience. However, the associated security risks and privacy concerns can prevent such data disclosures to some extent. Such privacy-benefit trade-offs are worth more in-depth research.

Furthermore, there is also research work investigating how to best store and manage personal data. For instance, Verbrugge et al.~\cite{Verbrugge-S2021} examined the possibility for a ``personal data vault society" and the steps necessary to realise this vision. The alternative being explored is an ecosystem of personal data spaces, in which every individual has their own personal data vault or data pod. A prerequisite for the effective implementation of a personal data vault-based ecosystem is the identification of one or more actors prepared to assume the role of the data pod provider(s). This position does not exist in the current ecosystem based on walled gardens. They gained an understanding of the trade-offs that end users, applications, and service providers face when selecting whether to participate in a personal data vault-based digital ecosystem. They briefly discussed the technological possibility of an individual data vault implementation based on Solid\footnote{\url{https://solidproject.org/}}. Fallatah et al.~\cite{Fallatah-KU2023} reviewed existing work on personal data stores (PDS), which allow individuals to store, control, and manage their personal data. They suggested that before PDS can be successfully and extensively used, the supporting platforms must overcome numerous obstacles and difficulties. They anticipated that these advantages will benefit individuals, organisations, and society. While PDS platforms are primarily concerned with assisting individuals to recover ownership over their personal data, businesses would welcome access to cleaner, more prosperous, and more securer personal data. They also argued that one of the technical barriers is the data flow management between systems and applications, and automatic and semi-automatic validation of processes performed by PDS platforms. In addition, another way to consolidate the understanding of privacy and personal data collection/sharing is to develop graphical models. Considering the positive correlation between the amount of personal data online and the number of intentional and unintentional privacy violations, data privacy is becoming one of the online users' primary concerns. Due to the lack of high-level modelling notations and the limited availability of low-level enforcement approaches for privacy aspects, not all privacy requirements associated with a given context can be adequately addressed. Labda et al.~\cite{Labda-W2014} proposed a novel extension to the Business Process Model and Notation (BPMN) to address privacy concerns. Dang et al.~\cite{Dang-TT2019} proposed a graphical model for representing and protecting the privacy of users on online social networks. Their model has two major components: a privacy object and a controller. The role of the privacy object is to capture the privacy characteristics of online social networks' actions. The part of the privacy controller is to ensure that end users can safely interact with the privacy object. Although the privacy controller is efficient for configuring privacy settings on social networks, its usability still needs evaluating.

Cyberthreats (circumstances or events with the potential to negatively impact organisational operations and assets, individuals, other organisations, or entire nations through an information system via unauthorised access, destruction, disclosure, or modification of information and denial of service) are increasing as a result of trends such as the proliferation of IoT devices. The efficient analysis of cyber threats is essential for risk assessment, cyber situational awareness, and security countermeasures. Still, it is contingent on sharing threat intelligence with context and rich semantics. More recently, a graphical model proposed by Lu and Li~\cite{lu2022data} can evaluate personal data flows from ``me'' (the user) and values flowing back to ``me'' to help inform the user about privacy-benefit trade-offs. This novel data flow-based graphical model can help individuals achieve a better understanding and management of the balance between privacy and benefits in the cyber-physical world. By analysing different types of edges in the proposed graphical model, they illustrated how their model can be used to identify privacy issues and benefits via analysis and visualisation of personal data and value flows. More specifically, different privacy issues can be represented and detected by different topological patterns involving one or more data flows.

\section{Methodology}
\label{sec:methodology}

In this work, we propose constructing possible flows of personal data through the analysis of an online travel service provider's privacy policy. By visually representing a personal data flow graph derived from the privacy policy, we intend to reveal some potentially overlooked details of personal data sharing activities of consumers of the online service provider.

We decided to use the privacy policy of Booking.com\cite{bookingcom_privacy}as a case study based on the following reasons: 1) Booking.com has the highest revenue globally within the online travel market and is the largest online travel agency by booking volume\footnote{\url{https://www.researchandmarkets.com/reports/5330849/global-online-travel-market-2022}}. 2) Booking.com provides a wide range of features and has a close link with many other subsidiaries of its parent company, Booking Holding Inc., therefore being a good case for understanding how personal data are shared between multiple parties. 3) Booking.com deals with their customers' personal data all the time and in large volume due to the nature of its business model. This requires its privacy policy to provide more details on how its consumers' personal data are collected, processed and shared.

To better facilitate the personal data flow mapping and visualisation, we adopted a simplified version of the graphical model proposed in~\cite{lu2022data} with the main aim of establishing the relationships between the following entities of different types: 
\begin{itemize}
\item `\emph{Person}' entities stand for natural people in the physical world

\item `\emph{Data}' entities refer to atomic (personal) data items about one or more person entities

\item `\emph{Service}' entities refer to different physical and online services that serve people for a specific purpose

\item `\emph{Organisation}' entities refer to organisations that relate to one or more services
\end{itemize}
We analysed the privacy policy from the perspective of how data entities flow from users of Booking.com to different data-consuming entities including Booking.com and other organisation entities. More specifically, we analysed the privacy policy from the following two main perspectives: 
\begin{enumerate}
\item \textbf{data collection} is about how Booking.com can implicitly and explicitly collect personal data from its customers, and how Booking.com may receive personal data about its customers from other sources indirectly (i.e., not from its customers directly);

\item \textbf{data sharing} is about how Booking.com shares personal data collected with third parties, including within Booking Holdings Inc.\ and its other subsidiaries, and with other third parties and online social media service providers.
\end{enumerate}

By manually noting down the relationships between different entities while going through the whole privacy policy, we were able to derive a graphical representation of possible personal data flows and a visualisation of the graph. It is worth noting that the graph presented in this paper is a simplified version, which is based on the assumption that the booker, referring to an individual who arranges a travel booking, is also the sole traveller. It is important to acknowledge that the personal data flows and the data flow graph can be more complicated when the booker is not a traveller or a member of a large group of travellers. For instance, when the booker is an employee of a company or travel agent or a friend/family member of the traveller(s) who does not participate in the booked travel, personal data of both the booker and of the traveller(s) will need to be shared with Booking.com. In such scenarios, personal data flows between the booker and the traveller(s) also need considering, which may also include the case that some traveller(s) may have an account on Booking.com while others do not have one. We consider such more complicated cases out of the scope of this study, and leave them as part of our future work. We also would like to highlight that, although the results presented in this paper are mainly based on the analysis of the privacy policy, we also made dummy bookings on Booking.com to recover and clarify some less apparent information to consolidate our understanding of personal data flows associated with a travel booking made using Booking.com.

\section{Results}
\label{sec:results}

Figure~\ref{fig:booking-data-economy} shows the reconstructed personal data flow graph through the analysis of the privacy policy of Booking.com in terms of personal data shared by consumers of Booking.com. As indicated by the green arrows at top of the graph, the personal data flows from the left to the right side demonstrate how Booking.com can collect their customers' personal data, what types of personal data can be collected, and to what extent Booking.com shares collected personal data with other third parties and for what purposes. In addition, we use `Challenge 1', `Challenge 2' and `Challenge 3' in Figure~\ref{fig:booking-data-economy} to represent three main challenges that we have identified, when identifying personal data flows based on analysis of a privacy policy.

\subsection{How Booking.com Collects Personal Data}

Based on the analysis of Booking.com's privacy policy, we investigated the scale and the scope of Booking.com's personal data collection and sharing activities. As illustrated in Figure~\ref{fig:booking-data-economy}, Booking.com can collect their customers' personal data from various sources in several ways. We categorise them into two groups based on the original data controller (the organisation who collects personal data in the first place): \emph{direct data collection} and \emph{indirect data collection}. It is worth noting that all personal data types listed in \emph{Data Boxes A, B, C, D} in Figure~\ref{fig:booking-data-economy} are extracted from examples in the privacy policy. However, we acknowledge that it is not an exhaustive list of the personal data types Booking.com may collect. We add three dots at the bottom of \emph{Data Box D} to indicate such incompleteness and also consider this as one of the challenges (i.e., Challenge~1 in Figure~\ref{fig:booking-data-economy}), which deserves further studies. We provide more details throughout the rest of this section.

\subsubsection{Direct Data Collection}
\label{subsubsec:direct_data_sharing}

Direct data collection refers to the case that Booking.com collects some personal data directly from their consumers. We further identified two approaches to direct data collection. 

\begin{figure}[!h]
\centering
\includegraphics[angle=90,height=.9\textheight]{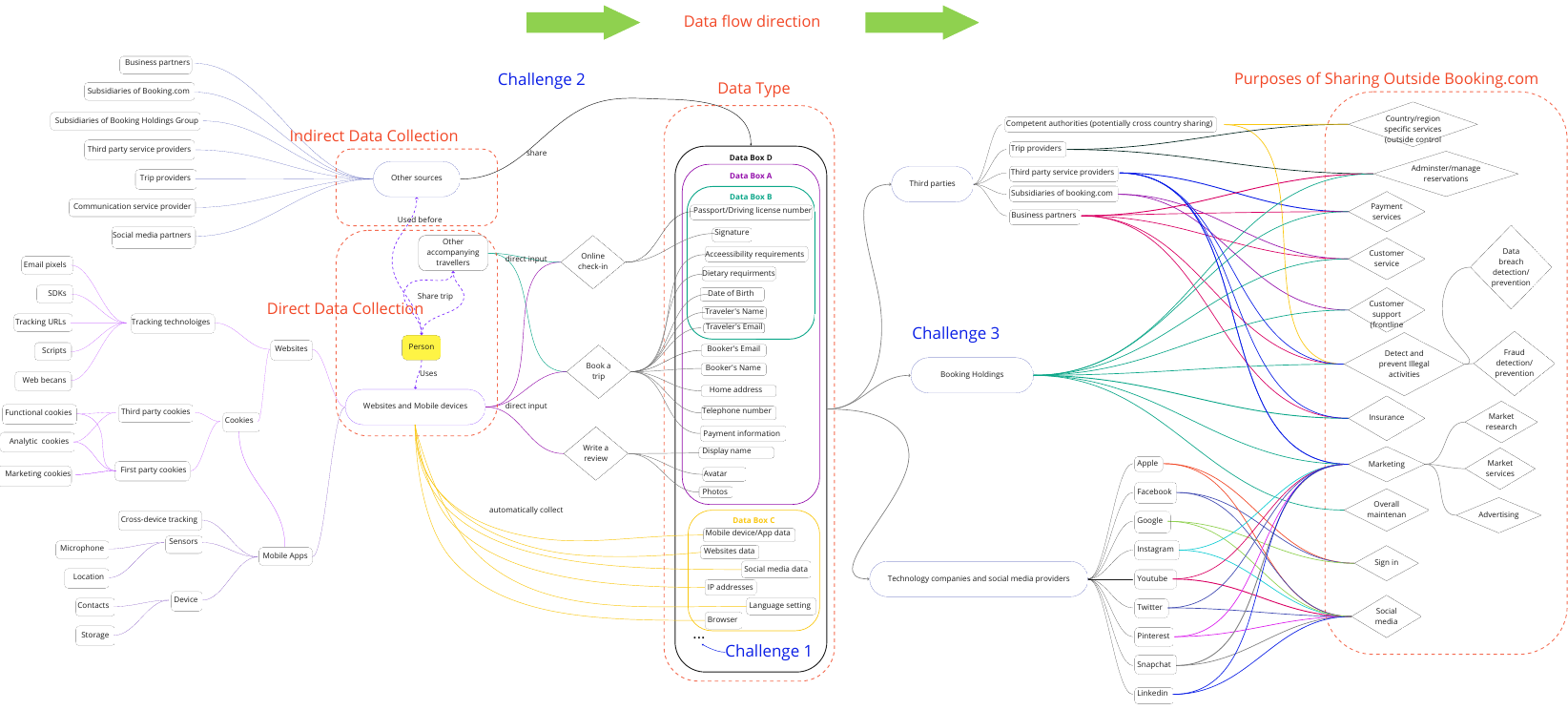}
\caption{Booking.com Personal Data Flows Diagram Extracted from its privacy policy}
\label{fig:booking-data-economy}
\end{figure}

\begin{itemize}
\item \emph{Explicit direct data collection} means that a person (the traveller or their assistant/helper) provides personal data to Booking.com directly via its website or mobile app as illustrated in Figure~\ref{fig:booking-data-economy}. Data Box A contains personal data that could be collected specifically for the booking purposes such as the booker's name, telephone number, and email address. Data Box B includes information about the traveller such as their name, date of birth, email address, and dietary requirements. 
Here we use two examples to better explain this. In addition, we present an example to illustrate typical types of personal data that can be collected for a range of selected online activities as shown in Table~\ref{tab:activity_data_type}.

\begin{table}[!htb]
\centering
\caption{Illustration of the typical types of personal data collected for different online activities}
\label{tab:activity_data_type}
\begin{tabularx}{\linewidth}{c X}
\toprule
\textbf{Online activity} & \textbf{Typical type(s) of personal data collected}\\
\midrule
\texttt{Online check-in} & name, date of birth, nationality, date of birth, place of birth, passport/driving license number, signature\\
\midrule
\texttt{Booking a trip} & name, date of birth, email address, home address, telephone number, payment information, accessibility requirements, dietary requirements\\
\midrule
\texttt{Writing a review} & display name, avatar, the profile image and photo(s) to support the review\\
\bottomrule
\end{tabularx}
\end{table}

\emph{Example 1:} 
If a booker is a traveller, Data Box A and Data Box B will overlap, indicating that the same personal data will be collected via Booking.com's website or mobile app for online booking activities such as online check-in, booking a trip and writing a review.
    
\emph{Example 2:}
If a booker is not a traveller but makes a booking for another person. For instance, a sectary makes a booking for its CEO. The sectary's personal data as represented in Data Box A will be collected for booking purposes, and the CEO's personal data as illustrated in Data Box B will be provided by the sectary to Booking.com. 

\item \emph{Implicit direct data collection} means that, without explicit data input from a user, Booking.com automatically collects some personal data, which is occurring simultaneously while a user is using Booking.com's website or mobile app. As depicted in \emph{Data Box C} in Figure~\ref{fig:booking-data-economy}, such personal data could include behavioural data of the user when using a mobile device, the website or the mobile app, the user's social media data, IP addresses, and language settings on the devices, which can be automatically collected using different technologies such as web tracking technologies, web cookies, device sensors, and cross-device tracking technologies. As mentioned in Section~\ref{sec:literature-review-and-concepts}, substantial work has been done by Jin et al.~\cite{Jin-H2018} to extract and visualise mobile apps' data flows\footnote{\url{http://android-network-tracing.herokuapp.com/}}.
\end{itemize}

\subsubsection{Indirect data collection}
\label{subsubsec:indirect_data_sharing}

Indirect data collection refers to personal data that are not collected by Booking.com directly due to a booker's use of its website or mobile app but shared with Booking.com by other parties such as third-party service providers, associated business partners, and other subsidiaries of Booking Holdings Inc.

\subsubsection{Remarks}

Please note that these parties' privacy policies could have explicitly allowed sharing of their consumers' personal data with Booking.com. Thereby, as shown in \emph{Data Box D} in Figure~\ref{fig:booking-data-economy}, a user's personal data could be shared with Booking.com and be used together with other personal data directly collected by Booking.com to better serve the person. Here, we would like to highlight that \emph{Data Box D} may contain a much more comprehensive range of personal data, since Booking.com can still gain access to many personal data of their customers indirectly from other third parties. As stated in Booking.com's privacy policy, it is worth noting that how and what other third-party organisations share personal data with Booking.com depend on their business needs and privacy policies. In other words, it is impossible to get more insights about such personal data without analysing privacy policies from other third-party organisations, which is needed to recover the full scale of indirect data collection. We consider this as another challenge (i.e., Challenge~2 in Figure~\ref{fig:booking-data-economy}) for future research.

\subsection{How Booking.com Shares Personal Data}
\label{subsec:how-booking.com-share-data-with-other-parties}

Booking.com claims to collaborate with various organisations to provide more satisfactory services to their users and to serve other legitimate purposes, which often require sharing personal data between such collaborative organisations. By analysing the privacy policy of Booking.com, we have identified three main destinations for data sharing: Booking Holdings Inc.\ and its subsidiaries, third-party organisations, technology companies and social media providers. 

\subsubsection{Third-Party Organisations}
\label{subsubsec:third_parties}

As listed in Figure~\ref{fig:booking-data-economy}, Booking.com's privacy policy mentions several types of third parties. Each type has its specific purpose of utilising personal data shared by Booking.com to fulfil users' booking activities. For instance, a) authorities:  personal data can be shared with national or local authorities for legal requirements or legal compliance purposes; b) business partners: Booking.com outsources its administration/reservation services and customer services to a business partner, which would require sharing some personal data to facilitate such outsourced services; c) payment service providers: Booking.com can share their customers' personal data with third-party payment service and insurance providers. The privacy policy does not provide a detailed list of companies working with Booking.com. Therefore, the full scope of Booking.com's data-sharing behaviours cannot be fully recovered by analysing its privacy policy alone.

\subsubsection{Booking Holdings Inc.\ and Its Subsidiaries}
\label{subsubsec:booking_holdings}

Being a subsidiary company of Booking Holdings Inc., Booking.com can share their users' personal data upward to its parent company. Booking Holdings Inc.\ provides travel-related services to people in more than 220 countries and territories through many subsidiaries. Besides Booking.com, some well-known brands include Priceline, Agoda, Rentalcars.com, KAYAK, OpenTable, Rocketmiles, Fareharbor, HotelsCombined, Cheapflights  and Momondo~\footnote{\url{https://www.bookingholdings.com/about/factsheet/}}. These subsidiaries of Booking Holdings Inc.\ offer some essential services to each other, and personal data collected by Booking.com may be further spread to other subsidiaries in order to provide more sophisticated combinations of services, as shown in Figure~\ref{fig:booking-data-economy}, including payment services, admin/manage reservations, customer support, illegal activity detection and prevention.

\subsubsection{Technology Companies and Social Media Providers}

Booking.com can share their users' personal data with technology companies and social media service providers in exchange for their services or to provide extra customer benefits. For example, Booking.com allows customers to sign in using their Apple, Facebook or Google credentials. This way, some additional personal data from that account may be shared with Booking.com, e.g., email address, age, and profile picture~\cite{bookingcom_privacy}. In addition, using social media features such as integrating social media plugins into Booking.com's website or mobile app and social media messaging services can lead to exchanges of personal data between Booking.com and social media service providers (e.g., Instagram, YouTube, and Twitter). As illustrated in Figure~\ref{fig:booking-data-economy}, several social media providers, including Instagram, YouTube, Twitter, Pinterest and Snapchat, are working with Booking.com in such a way. Moreover, another important purpose of such personal data sharing is to conduct market research and to provide more personalised market and advertising services. 

\subsubsection{Remarks}

Booking.com's privacy policy does not contain any information about which type of personal data are shared with which other organisations in detail, preventing us from achieving a full understanding of the landscape of personal data collection and sharing. Thereby, we envisage that another challenge (i.e., Figure~\ref{fig:booking-data-economy} Challenge~3) for future work is to conduct more work to study cross-organisational data sharing to consolidate our understanding.

\section{Further Discussions and Conclusion}
\label{sec:discussion-and-conclusion}

The paper presents a case study to understand how Booking.com collects and shares personal data based on analysis of its privacy policy. By producing a personal data flow graph as a visual aid, we were able to reveal how Booking.com can collect personal data and shares such data with other organisations. Although our work focuses on Booking.com as a case study, the following lessons learnt are likely true for many other online services regarding the challenges of refining privacy policies to reflect the landscape of personal data collection and sharing: 
\begin{itemize}
\item The lack of a comprehensive description of the types of personal data that could be collected directly or indirectly
\item An incomplete description of how and to what extent other organisations can share personal data with online service providers
\item An unclear description of how and to what extend online services can share their customers' personal data with other third parties;
\item An unclear disclosure on how personal data collected are used. The lack of clarity and transparency makes it difficult for users to understand the full extent of personal data collection and sharing and how their personal data may be used, therefore subsequently harming their confidence in continuously accepting the business-centric approach to personal data management of online services' users~\cite{wieringa2021data}.
\end{itemize}

Furthermore, this study has the following limitations, and we intend to address these in our future work. 

\begin{itemize}
\item We only considered using the privacy policy as the only data source to derive the personal data flows, which is not enough to establish a comprehensive understanding of how personal data is collected and shared.

\item In order to not over-complicate the personal data flow graph, some more fine-grained details of data flows are not presented in this paper. For instance, not all data listed in Figure~\ref{fig:booking-data-economy} will be shared with social media providers, and the type of personal data that can be shared with a social media provider will depend on the purpose of using the related social media account.
 
\item A study involving real users should be conducted to get direct feedback from end users to evaluate the usefulness and effectiveness of our approach. 
\end{itemize}

Last but not the least, we consider our work as the basis to establish a more general approach to automating extraction of personal data flows for any given online services. We hope that this case study can be a stepping stone to elicit more follow-up work on privacy policy analysis, personal data flows, and related graphical modelling and ontological research.


To further expand the discussion for future research directions, this study suggests that such large scale with less controlled nature of user data sharing has been legitimised through privacy policy and acclaimed compliance with legislation such as GDPR. To some extent, the root cause of the loss of data sharing control could be visualised in data breaches and fraud in such data flow diagram. Arguably, the root cause for such data privacy and security issue is due to the centralised data collection of personal data by service providers. So far, such firm-centric control of the personal data model has fuelled the data economy~\cite{economist_worlds}. Wieringa et al.~\cite{wieringa2021data} has contrasted the firm-centric model with a person-centric model and concluded that there is scope for shifting data control to individuals; allowing customers to decide and originate data sharing to limited firms after assessing the benefits of doing so.

Following this school of thought, we envisage a foundational shift in the data economy, where individuals are empowered and lifted as data controllers with adequate technological and contractual capability. In such a scenario, an individual will be able to act just like a firm to participate in the data exchange and gain a share of the data economy. Organisations that offer such capability are gaining traction worldwide, namely SOLID\footnote{url{https://solid.mit.edu/}}, Digi.me\footnote{\url{https://digi.me/}}, hubofallthings (HAT)\footnote{\url{https://www.hubofallthings.com/}}.

Imagine an individual replacing any or even every actor of the booking.com data sharing network by sending service requests (signals) to providers without emitting personal data to them (i.e. an API endpoint for firms to curate their services on a personal server). As individuals can produce demand in a real-time and contextualised manner, service providers will be more incentivised to provide more personalised service to match customers’ direct signalling of service requisition directly.

Moreover, such privacy-preserving models could further create a fundamental power shift of data ``ownership'' rights from firms to individuals. For example, SOLID and HAT put individuals back in legal ownership and control of their data. In this case, service providers must treat individuals as equal peers, i.e., as suppliers or service providers. Consequently, data aggregators like Booking Holdings may lose their competitive advantage as their data aggregation power may be stripped. Regulatory-wise, individuals acting as data ``controllers'' and ``processors'' are exempted from the current regulatory landscape Information Commissioner’s Office, 2014. However, there are potential consequences in that innovative firms may also inherit the exemption and gain immunity from data protection compliance if they only execute service rendering directly from individual servers.


\section*{Acknowledgements}

The authors were supported by the Engineering and Physical Sciences Research Council, UK Research and Innovation (UKRI), as part the project ``PriVELT: PRIvacy-aware personal data management and Value Enhancement for Leisure Travellers'', under the grant numbers EP/R033749/1 and EP/R033609/1.

\bibliographystyle{splncs04}
\bibliography{main}

\end{document}